\documentclass{aastex}
\usepackage{spr-astr-addons}

\RequirePackage{color}

\begin{document}

\title{High Frequency Cutoff and {\bf{Change}} of Radio Emission Mechanism in Pulsars}
%% Running heads
\shorttitle{High Frequency Cutoff in Pulsars} \shortauthors{V. M.
Kontorovich and A. B. Flanchik}

\author{V. M. Kontorovich\altaffilmark{1,2}}
\and \author{A. B. Flanchik\altaffilmark{1}}

\altaffiltext{1}{Institute of Radio Astronomy of National Academy
of Sciences of Ukraine, 4 Chervonopraporna Str., Kharkov 61002,
Ukraine.} \email{\vkont1001@yahoo.com}

\altaffiltext{2}{Karazin Kharkov National University, 4 Svobody
Sq., Kharkov 61077, Ukraine.}

\begin{abstract}
Pulsars are the fast rotating neutron stars with strong magnetic
field, that emit over a wide frequency range. In spite of the
efforts during 40 years after the discovery of pulsars, the
mechanism of their radio emission remains to be unknown so far. We
propose a new approach to solving this problem for a subset of
pulsars with a high-frequency cutoff of the spectrum from the
Pushchino catalogue (the "Pushchino" sample). We provide a
theoretical explanation of the observed dependence of the
high-frequency cutoff from the pulsar period. The dependence of
the cutoff position from the magnetic field is predicted. This
explanation is based on a new mechanism for electron  radio
emission in pulsars. Namely, radiation occurs in the inner (polar)
gap, when electrons are accelerated in the electric field that is
increasing from zero level at the star surface. In this case
acceleration of electrons passes through a maximum and goes to
zero when the electron velocity approaches the speed of light. All
the radiated power is located within the radio frequency band. The
averaging of intensity radiation over the polar cap, with some
natural assumptions of the coherence of the radiation, leads to
the observed spectra. It also leads to an acceptable estimate of
the power of radio emission.
\end{abstract}

%% Keywords

\section{Introduction}%\label{s:intro}

Pulsars are magnetized neutron stars that have a magnetosphere
filled with an electron-positron plasma of about the GJ density
\citep{b49, b39, b3}. New discoveries of double pulsar system
\citep{b27} and intermittent pulsars \citep{b25, b29, b6} give the
direct observational support to that idea. It is thought that this
plasma in the region of open magnetic field lines over the
magnetic polar cap is generated by particles (through gamma quanta
production) accelerating in a gap under the magnetosphere
\citep{b51, b47, b1, b2}. The acceleration of electrons occurs in
the gap in the electric field that is longitudinal with respect to
the magnetic field and induced by the rotation of the magnetized
star. Directed coherent electromagnetic radiation of relativistic
particles from the region of open lines creates the beacon effect
that results in the pulses observed (the most popular
explanation).

In explanation of radio emission of pulsars (see reviews
\citep{b36, b38} and addition references in \citep{b37, b20, b4})
the instabilities of plasma flow, beam instabilities and similar
effects in the magnetospheric plasma\footnote{Note apart the
plasma-beam (see as example \citep{b52, b55}), also the cyclotron,
drift, modulation instabilities, Zakharov's wave collapse and
magnetic reconnection for GP, low-frequency "tails" of the
synchrotron, Cherenkov, Doppler and curvature radiation in the
relativistic electron-positron plasma. } have been discussed.
Apparently, various mechanisms of radio emission are actually
realized and may in certain circumstances succeed each other.

We show in this paper that for the observed pulsar radio emission
a coherent radiation produced in a polar gap may be responsible,
at least for pulsars of the Pushchino sample\footnote{The sample
is based on Pushchino catalogue \citep{b31,b32}. Catalogue gathers
simultaneous and compiled radio spectra for 340 pulsars (on more
than three frequencies) with more than 120 references on
measurements on different instruments including Bonn 100m and
Ukrainian decameter radio telescopes. The correlation between
frequency maximum, cutoff frequency and period have been found for
some part of the spectra \citep{b33}. We don't touch here the more
high frequency values in catalogue \citep{b40} with new results
that ask for separate investigation, see also discussion in
\citep{b48}.} with cutoff in the radio spectrum. The relationships
we have received we will also use for the pulsar in the Crab
Nebula, given its peculiarities. The radiation has been emitted
with the acceleration of electrons in the gap. It is quite
essential that the accelerating longitudinal electric field in the
gap slowly increases from zero at rising off the star surface.
This mechanism has never been discussed earlier for pulsars.

In the well-known Ruderman-Sutherland model the strong electric
field, non vanishing at the star surface, accelerates electrons so
quickly that their radiation due to acceleration in the gap fully
comes to the hard energy (X-ray) region with no radio emission.
Only in an electric field slow rising from zero level on the star
surface the radiation of accelerating electrons comes to the radio
band.

\begin{figure}
\includegraphics[angle=0,scale=1]{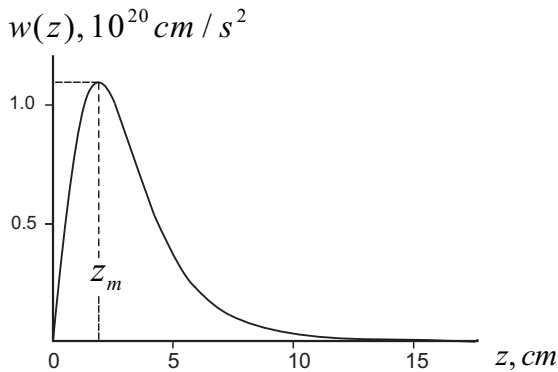}
 \caption{The dependence of the
acceleration on the altitude $z$ on the star surface}
% \label{fig:1}
\end{figure}

Resulting radiation is limited by a cutoff frequency found in this
study. It coincides with the high-frequency cutoff \citep{b33,
b36} in the pulsar spectra of the Pushchino sample. Such frequency
limitation is due to the fact that the electron acceleration in
the electric field, vanishing on the star surface, passes through
a maximum and decreases as the electron velocity approaches the
relativistic limit.

There is the all-new specification of our point of view and the
main result of this work.

\section{Acceleration of electrons in the speed up process}
We start from the equation for the electron Lorentz factor
$\Gamma(z)$ \citep{b28}.

\begin{equation}\label{eq1}
d\Gamma \left( {z} \right)/dz = eE\left( {z} \right)/mc^{2},
\end{equation}

\noindent where $z$ is the altitude of the electron above the star
surface. The equation describes the change of electron energy in a
nonuniform electric field $ E(z)= E_z (z)$. (Only component that
is parallel to the strong magnetic field $\bf{B}$ directed on z
axes is essential). Energy losses (neither by curvature radiation
nor by inverse Compton scattering) are insignificant for relevant
for us values of Lorentz-factors $\Gamma$ of order of some units.
We also ignore, within the polar cap, the deviation of the
magnetic field lines from orthogonality to the surface of the
pulsar.

The low-frequency radiation occurs at the small altitudes when the
accelerating electric field $E(z)$ rises from zero level on the
surface of the pulsar $z=0$. The vanishing of the electric field
on the star surface is related to a small electron work function
\citep{b18} of the surface. At altitudes $z < < h$, where
\textit{h} is the height of the gap, the field
$$
E\left( {z} \right) = E_{0} z/h
$$
increases linearly \citep{ b1, b14, b8, b2} with the altitude
\textit{z}. The velocity $V$ and the acceleration $w \equiv \ddot
{z}$, expressed in terms of Lorentz factor $\Gamma $, are equal to
$V = c\sqrt {\Gamma ^{2} - 1} /\Gamma $, and
$$
w\left( {z} \right) = eE\left( {z} \right)/m\Gamma ^{3}=
c^{2}\Gamma ^{ - 3}{\Gamma}'
$$
where ${\Gamma} ' \equiv d\Gamma /dz$. In a linear
field we have
$$
\Gamma \left( {z} \right) = \Gamma _{0} + az^{2}, a = eE_{0}
/2mc^{2}h.
$$
The acceleration increases from zero when $z = 0$, passes through
a maximum (Fig.1) when
$$
\Gamma _{m} = 6/5, \, \, V_{m} = c\sqrt {11} /6
$$
(not dependent on the field) and
$$
z_{m}^{2} = 2mc^{2}h/5eE_{0}
$$
(for $\Gamma _{0} = 1$) and tends to zero at approaching of
electrons to relativistic velocities. The value of particle
acceleration at the maximum is
$$
w_{m} = \left( {5/6}
\right)^{3}c\sqrt {2eE_{0}/5mh}.
$$
The height of the gap falls from these relations when used
henceforward the field of the form \citep{b44}
$$
\quad E_{0} \sim \Omega \cdot Bh/c .
$$
Physically, this estimate is quite obvious since on the scale of
the gap the characteristic velocity due to rotation is $\Omega
\cdot h$. We omit here a factor of order of unity that will be
partly considered below, which contains the dependence of $E_{0} $
on the position on the polar cap. We do not discuss here the
influence of general relativity effects and dependence on the
angle between the axis of rotation and magnetic axis of the pulsar
\citep{b2}, which does not affect the estimates. (The exception is
PSR B0531+21 which is close to an orthogonal rotator.) The
estimate
$$
z_{m} \approx \sqrt {\left( {P/1{\kern 1pt} {\kern 1pt} s} \right)
\cdot \left( {10^{12}G/B} \right)} \cdot 1{\kern 1pt} {\kern
1pt}cm
$$
confirms the legality of the conditions $z << h$ which we use, as
$h \sim 10^4 cm $ for normal pulsars.

\section{Radiation at acceleration in the gap}

For the radiation field of accelerated electrons, we proceed from
the retarded Lienard-Wiechert potentials \citep{b17, b28}. In the
problem considered the particle acceleration {\bf{w}} is directed
along its velocity {\bf{V}}. Then for the Fourier component of the
wave magnetic field we have (at large distances from the radiating
electron)

\begin{equation}\label{eq2}
{\it\bf{H}}_{\omega}  = \frac{{e}}{{c^{2}R_{0}} }e^{ikR_{0}}
\int\limits_{ - \infty} ^{\infty}  {\frac{{\left[ {{\it\bf
{w}}\left( {t} \right),{\it\bf{n}}} \right]}}{{\left( {1 -
{\it\bf{n}}{\it\bf{V}}\left( {t} \right)/c}
\right)^{2}}}e^{i\left( {\omega \,t - {\it\bf{k}}{\kern 1pt}
{\it\bf{r}}\left( {t} \right)} \right)}} dt.
\end{equation}

\noindent {\bf{Here}} $t = t\left( {z} \right) =
\int\limits_{0}^{z} {} d{z}'/V\,\left( {{z}'} \right)$,
${\it\bf{r}}\left( {t} \right)$ is the electron radius vector,
$R_{0} $ is the distance from the origin on the star surface $z=0$
to the field observation point, ${\it\bf {n}} = {\it\bf {k}}/k$.

A connection of the time $t$ with the electron coordinate $z$ is
given by the integral

\begin{equation}\label{eq3}%2a
t\left( {z} \right) = \frac{{1}}{{c}}\int\limits_{0}^{z}
{\frac{{\Gamma \left( {{z}'} \right)d{z}'}}{{\sqrt {\Gamma
^{2}\left( {{z}'} \right) - 1}} }},
\end{equation}\label{eq1}

\noindent where $\Gamma(z)$ is governed by (1).

To eliminate the logarithmic divergence at zero one should take
into account in the difference $\Gamma - 1$ that the initial
velocity $V\left( {0} \right) = V_{T} \ne 0$ and therefor
$\Gamma_{0} \approx 1 + V_{T}^{2} /2c^{2}$.
 Here $V_{T} << c $ is the thermal velocity of electrons
 on the surface of the pulsar polar cap. It is convenient to
 represent the Fourier component ${\it\bf{H}}_{\omega } $,
 which determines the emission spectrum, as an integral over coordinate:

\[
{\it\bf{H}}_{\omega}  = \frac{{e}}{{c^{2}R_{0}} }e^{ikR_{0}}
\left[ {{\it\bf{l}},{\it\bf{n}}} \right]L_{\omega},\] where $\quad
{\it\bf{l}} = {\it\bf{V}}/V,\;{\it\bf{n}} =
{\it\bf{k}}/k,\;{\it\bf{l}} \cdot {\it\bf{n}} = cos\theta $ and

\begin{equation}\label{eq4}
L_{\omega}  = \int\limits_{0}^{h} {\frac{{w\left( {z}
\right)}}{{V\left( {z} \right)\left( {1 - {\textstyle{{V\left( {z}
\right)} \over {c}}}cos\theta} \right)^{2}}}e^{i\omega \,\left(
{t\left( {z} \right){\kern 1pt} - \frac{{z}}{{c}}cos\theta}
\right)}} dz.
\end{equation}

Accordingly, we have for the spectral and angular radiation
intensity density $d\varepsilon _{n\omega}/ d\omega do$
interesting us \citep{b17, b28}

\begin{equation}\label{eq5}%3
d\varepsilon _{n\omega}  = \frac{{e^{2}}}{{4\pi
^{2}c^{3}}}sin^{2}\theta \cdot \left| {L_{\omega} }
\right|^{2}d\omega do.
\end{equation}

\noindent Here $d\omega$ is the interval of frequencies, $do$ is
the interval of solid angles. It is seen from the integral
$L_{\omega}  $ that due to rapid oscillations the field decreases
exponentially for $\omega > \omega _{cf} $, where the cutoff
frequency is $\omega _{cf} \approx \pi \sqrt {2eE_{0} /mh} $
(Fig.2). This estimate can be obtained numerically from Eq.(5) and
independently from the physical consideration, supposing that the
electron movement becomes relativistic: $e\int\limits_{0}^{z_{cf}}
{E\left( {z} \right)dz} = mc^{2}$ and $\omega _{cf} = 2\pi {\kern
1pt} c/z_{cf} $. We obtain $\Gamma _{cf} = 2$, $V_{cf} = \sqrt {3}
c/2$, $z_{cf}^{2} = 2mc^{2}h/eE_{0} $=$5z_{m}^{2} $. The
coefficient $a = 1/\lambda^2_{cf} $, where $\lambda _{cf} $ is the
wavelength corresponding to the cutoff.

\begin{figure}
\includegraphics[angle=0, scale=0.9]{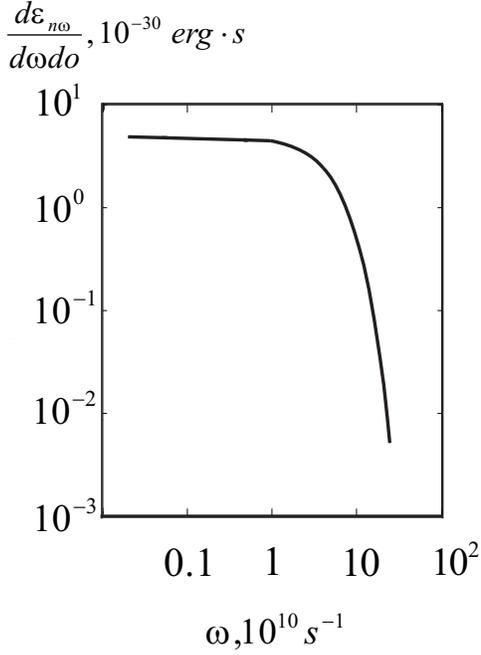}
 \caption{The emission spectrum of a particle at acceleration in the
electric field linearly increasing with altitude z above the
surface of the star (the numerical calculation of (5) for
$\theta=\pi/8, \,B = 10^{12} G, \,P = 1 s$)}
\end{figure}

To move to the average spectra we take into account the dependence
of the field from its position on the polar cap ( cf. \citep{b8})
of the form:

\begin{equation}\label{eq6}
E_{0} \left( {r} \right) = E_{max} \left( {1 -
\frac{{r^{2}}}{{R_{PC}^{2} }}} \right), \quad R_{PC} \approx
R\sqrt {\frac{{\Omega R}}{{c}}}
\end{equation}

\noindent ($R $ is the star radius). Assuming for the maximum
field at the center of the polar cap $E_{max} /B = \Omega \cdot
h/c$, we obtain the value of the cutoff frequency $\omega _{cf}
\left( {0} \right) = \pi \sqrt {2e\Omega B/mc} $, which is
independent of the height gap. The cutoff frequency dependence on
the pulsar magnetic field is shown in Fig.3.

\begin{figure}
\includegraphics[angle=0,scale=0.9]{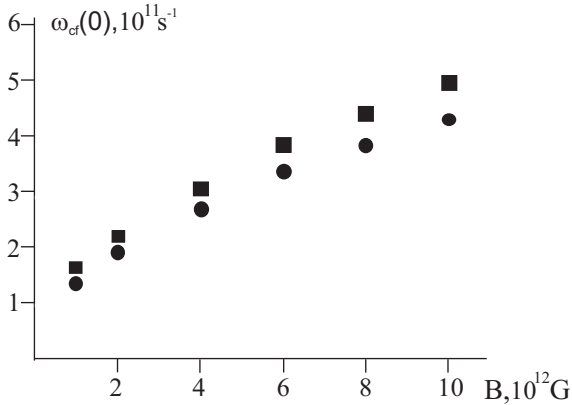}
 \caption{The dependence of cutoff frequency on the magnetic field
 for $\theta=\pi/8, P=0,07 s$. Squares
correspond to $\omega _{cf} \left( {0} \right)$, circles -- to
numerical integration of $\L_\omega $}
% \label{fig:3}
\end{figure}

At this frequency, the emission spectrum for the discussed
mechanism must cutoff and be replaced by the next highest, forming
a kink. The cutoff frequencies for a sample of normal pulsars had
been found in \citep{b33}. The dependence on the period for $B = 2
\cdot10^{12}G$ coincides for $\hat{\nu} _{cf}=V _{cf}/2\pi z_{cf}$

\begin{equation}\label{eq7}
\hat\nu _{cf} \approx \sqrt {2} \cdot 10^{9}Hz\sqrt {\left(
{\frac{{B}}{{2 \cdot 10^{12}G}}} \right) \cdot \left(
{\frac{{1\;s}}{{P}}} \right)},
\end{equation}

\begin{center}(this theory)
\end{center}
$$
\quad \tilde\nu _{cf} \approx 1.4 \cdot 10^{9}Hz\;\left(
{\frac{{1\;s}}{{P}}} \right)^{0.46 \pm 0.18}
$$
\begin{center}\citep{b33}.
\end{center}
(The {\bf{tilde}} over frequency means the experimental value).
 The dependence $\hat{\nu} _{cf}$ on $B$ and other parameters,
 {(see as example \citep{b8})
determining the electric field in the gap, is available for
testing. It can also serve as a criterion for the correctness of
this description. Below we use the cutoff frequency $\nu _{cf}=c/z
_{cf}$ corresponding decrease of intensity radiation of a single
electron to $e$ times (see Fig.2). The actual frequency of the
cutoff may be less than the estimated cutoff frequency both due to
the fact that the actual accelerating field in the gap is less
than the adopted estimates and the cutoff corresponds to higher
values of gamma-factor.

\section{Average spectra}\label{s:ja}
Now we consider the emission spectrum of a large number of
electrons accelerated in the electric field of the gap. From the
intensity of a single particle (taking into account that in the
electric field, linearly increasing from zero on the star surface,
the acceleration maximum is proportional to the square root of
$E_0$)

\begin{equation}\label{eq8}
I\left( {r} \right) = \frac{{2e^{2}w^{2}\left( {r}
\right)}}{{3c^{3}}} \approx \frac{{e^{3}E_{0} \left( {r}
\right)}}{{mch}},
\end{equation}

\noindent and changing the spectrum in the Fig.2  by the
step-function, we turn to the spectral density for $\omega <
\omega _{cf} \left( {r} \right)$:

\begin{equation}\label{eq9}
I\left( {r,\omega}  \right)d\omega \approx I\left( {r}
\right)\frac{{d\omega }}{{\omega _{cf} \left( {r} \right)}},
\end{equation}

\noindent where $ \quad \omega _{cf} \left( {r} \right) = \pi
\sqrt {\frac{{2eE_{0} \left( {r} \right)}}{{mh}}}. $
\medskip

Accordingly, the frequency range of radiation of the single
particle has the form

\begin{equation}\label{eq10}
\omega ^{2} \le \omega _{cf}^{2} \left( {r} \right) = \frac{{2\pi
^{2}eE_{max}} }{{mh}}\left( {1 - \frac{{r^{2}}}{{R_{PC}^{2}} }}
\right),
\end{equation}

\noindent therefore for the spectrum (in an incoherent case) we
obtain

\[
I\left( {\omega}  \right) \propto 2\pi \int\limits_{0}^{b\left(
{\omega} \right)} {\;rdr\;} I\left( {r,\omega}  \right)N,
\]

\begin{equation}\label{eq11}
\quad b\left( {\omega}  \right) = R_{PC} \sqrt {1 - \frac{{\omega
^{2}}}{{\omega _{cf}^{2} \left( {0} \right)}}},
\end{equation}

\noindent where $N$ is the number of emitting particles measured
through the current across the polar cap. Integration over the
polar cap gives

\begin{eqnarray}
 I\left( {\omega}  \right) \propto
\int\limits_{0}^{b\left( {\omega} \right)} {rdr} \sqrt {1 -
\frac{{r^{2}}}{{R_{PC}^{2}} }} \propto 1 - \frac{{\omega
^{3}}}{{\omega _{cf}^{3} \left( {0} \right)}},\label{eq12}\\
 \omega_{cf} \left( {0} \right) \approx \pi \sqrt {\frac{{2eE_{max}
}}{{mh}}} , \,\, \omega \leq \omega _{cf}(0),\nonumber
\end{eqnarray}

\noindent i.e. the spectrum is flat and {\bf{has}} a cut off at
the frequency $\omega _{cf} \left( {0} \right)$.

\begin{figure}
\includegraphics[angle=0,scale=0.8]{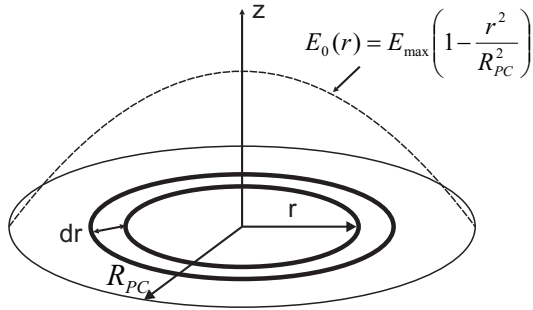}
 \caption{Formation of average spectrum of radio emission. The dotted
line shows the change of the longitudinal electric field with
{\bf{the}} distance from the magnetic axis to the boundary of
{\bf{pulsar}} polar cap}
% \label{fig:4}
\end{figure}

Let us now consider the coherence of the emission
\citep{b11,b10,b26, b12}. In the discussed mechanism only thin a
disk with $z<z_{cf}$ emits and the disk thickness is less then any
wavelengths emitted. Therefore, in our case there is no need to
raise additional assumptions about formation in the longitudinal
direction coherently emitting electron bunches, which represents
up to now a significant difficulty in all theories of pulsar radio
emission. However, we must {\bf{assume}} that in the disk plane a
fragmentation takes place to coherently emitting regions, just as
it is supposed in explaining the drift of subpulses \citep{b30}.
Here we restrict ourselves to the case when fragmentation occurs
in a region with a transverse dimension less than the length of
the radiated wave\footnote{In the case of a thin disk it is
possible, in principle, a coherent radiation of considerably
larger in transverse dimension regions, bounded by the scale
$\sqrt{R_0\lambda}$ \citep{b49}, which gives for the upper
boundary of power a significantly large value. When $R_0=R _{PC}$
it gives about a dozen of coherently emitting regions on the polar
cap, comparable with the number of observed ones in the drift of
subpulses \citep{b7,b49,b46}. See also \citep{b53}.}. In this
case, we find the upper limit of radiated power at a given
fragmentation. For this the total number of particles \textit{N}
is divided into coherent blocks (which are different for different
wavelengths $\lambda $ and depend on blocks position)

\begin{equation}\label{eq13}
N = N_{block} \cdot \frac{{2\pi} }{{\Sigma _{PC}}
}\int\limits_{0}^{R_{PC}} {rdr} N_{coh} \left( {r} \right).
\end{equation}

Inside of the blocks the intensity is proportional to the square
of the particle number $N_{coh} \left( {r} \right)$. The required
limitation \citep{b12} of block height $\bar {z}$ is provided by
the condition $\lambda > \bar {z} \approx z_{cf} $, where $z_{cf}
$ is the height to which the radiation mechanism acts. The blocks
are summed additively
\begin{equation}\label{eq14}
I\left( {\omega}  \right) \propto 2\pi \int\limits_{0}^{b\left(
{\omega} \right)} {\;rdr\;} I\left( {r,\omega}  \right)N_{block}
\left( {r} \right)N_{coh}^{2} \left( {r} \right).
\end{equation}

The condition $\lambda > \bar{z}$ for all emitted
waves is fulfilled automatically in our case that is very important
for realization a coherent emission mechanism. As for division of the
polar cap on coherent areas, at this stage we must restrict the discussion
only by assumptions.  Dividing polar cap on the regions of order lambda,
we obtain, in coherent case, the lower estimates for intensity of radiation
due to the minimum square of the number of particles in each region.

Obviously, the number of blocks equals to $N_{block} \approx
\left( {R_{PC} /\lambda}  \right)^{2}$, and the number of
particles in the block is $N_{coh} \approx \pi \lambda ^{2}\bar{n}
_{e}\bar{z} $.  Then for radiation spectrum we have

\begin{equation}\label{eq15}
I\left( {\omega}  \right) \propto 2\pi \int\limits_{0}^{b\left(
{\omega} \right)} {rdr\;} \frac{{1}}{{\lambda ^{2}}}\sqrt {E_{0}
\left( {r} \right)} \lambda ^{4}\bar {z}^{2} \propto
\end{equation}

\[\propto
\frac{{1}}{{\omega ^{2}}}\int\limits_{0}^{b\left( {\omega}
\right)} {\frac{{rdr}}{{\sqrt {1 - r^{2}/R_{PC}^{2}} } }} =
\frac{{1}}{{\omega ^{2}}}\left( {1 - \frac{{\omega} }{{\omega
{}_{cf}\left( {0} \right)}}} \right).
\]

\noindent where $b(\omega)$ is given by Eq. (11) and the frequency
$\omega _{cf} (0)$ is determined accordingly to Eq. (12). We have
obtained a power-law spectrum with spectral index 2 (Fig.5), which
is typical for the majority of pulsars. Let us note, that Eq.(12)
and (15) are only the asymptotics of the exact spectra. In reality
there is a low frequency break (or turnover) at the frequency
$\omega_{tr}\approx 0.1 \omega_{cf}$ \citep{b36}. Such spectrum
behaviour can be obtained in the discussed model\footnote{Note,
that there is a dynamical but not a dissipative reason (cf.
\citep{b45}) for a turnover.} and we have considered in detail the
physical meaning of that break and its position in the separate
paper \citep{b24}. Near the cutoff frequency the emission spectrum
can not be considered strictly as a power law. For a small number
of measured values it may be perceived as a kink. The spectral
index depends on transverse dimensions of coherent blocks, which
it seems also reveals themselves in the geometry of the subpulse
drift.

\begin{figure}
\includegraphics[angle=0,scale=0.8]{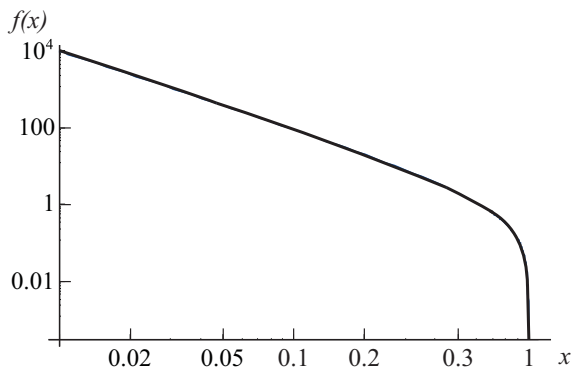}
 \caption{Function $f(x) = (1-x)/x^2, \, x= \omega/\omega
 _{cf}(0)$
 that defines the spectrum of coherent emission of electrons
accelerated in the vacuum gap}
% \label{fig:5}
\end{figure}

The coherence provides a reasonable estimate of the intensity of
radio emission. The rough estimate of the power of radio emission
has the form

\begin{equation}\label{eq16}
I_{R} \sim N_{block} N_{coh}^{2} I_{1} , \quad I_{1} \approx
\frac{{e^{3}E_{0}} }{{mch}}
\end{equation}

\noindent where $I_{1} $ is the radiation power of a single
particle at the maximum of acceleration. Here $N_{block} \approx
\left( {R_{PC} /\lambda _{max}} \right)^{2}$, where $\lambda
{}_{max}\sim 10^{2}cm$ is a wavelength corresponding to the
maximum in the spectrum of radio emission. Writing an estimate for
the number of particles in a coherent block in the form $N_{coh}
\approx \pi \lambda _{max}^{2} \bar {z}{\kern 1pt} \bar {n}_{e} $,
where $\bar {n}_{e} \sim n_{GJ} = \Omega B/\left( {2\pi ce}
\right)$ is the average density of particles near the surface, we
have $I_{R} \sim \left( {\pi R_{PC} \lambda _{max} \bar {z}{\kern
1pt} n_{GJ}} \right)^{2}I_{1} $, whence the estimate $I_{R} \sim
\lambda _{max}^{2} \Omega ^{3}R^{3}B^{2}/c^{2}$ results. For the
parameters $B = 4 \cdot 10^{12}G$, $P = 1\;s$ it leads to $I_{R}
\approx 10^{28}erg/s$, which agrees well with the data on the
radio luminosity of pulsars. For the fast rotating pulsar B0531+21
with high magnetic field this estimate reaches 10$^{32}$ erg/s
taking into account its proximity to the orthogonal rotator.

Note, that the assumption about change of radiation mechanisms in
the high frequency cutoff region allows in principle to explain
\citep{b22} the main pulse disappearance of PSR B0531+21 at
frequencies near 8 GHz \citep{b13}. Really, from our point of view
at lower frequencies the radiation at longitudinal acceleration of
subrelativistic electrons gives the principal contribution to the
main pulse. This emission vanishes at the cutoff frequency that
may lead to disappearance of the pulse. The radiation due to
low-frequency tail of narrow directed aberrational relativistic
mechanism remains. This anisotropic radiation does not fall to the
main pulse window but reveals itself in the interpulse if its line
of sight is more close to the magnetic axis.

Note also that the discussed mechanism of radio emission (for
which the place of radiation is definite and occupies a thin layer
near the surface of the star) may use for checking the new methods
to determine the generation location by a dispersion delay of the
signal in the magnetospere \citep{b15}. The accuracy of such
methods is limited now by the lack of the adequate knowledge of
magnetosphere properties.

\section{Conclusions}
We have found above that the radiation of accelerating electrons
in the electric field, slow increasing from zero on the star
surface, entirely comes to the radio spectral band. This is the
main result of the work.

The considered approach allows us explain the radio emission for
the pulsars from Pushchino sample, including the position of the
spectrum cutoff. As a result, we obtain a power-law spectrum,
which arise at averaging over the polar cap due to dependence of
the accelerating electric field on its position. The density
radiation in the gap is large without any assumptions about the
gap as a cavity (cf. \citep{b20}). It explains also the part of
the gamma-ray emission from pulsars and its correlation \citep{b5}
with giant pulses \citep{b21}.

The radiation with the linear acceleration was considered in
\citep{b41, b42}  for acceleration in a so strong electric field
on the star surface, when an electron reaches relativistic
velocities in a time shorter then the period of the wave emitted.
In this case the considered effects are absent.

The obtained results make it also possible to compare  the
theories of accelerating fields in the inner gap \citep{b14, b8}
(that is the foundation of all physics of pulsars) with
observations, transforming them from a "thing in itself" that is
not available to direct observations, in the "thing for us".

 \acknowledgments
{We are grateful to our colleagues from RI NAS of Ukraine and to
participants of PRAO-2011, JENAM-2011 and NS-2011 conferences for
discussions, to M.~Azbel', O.~Ulyanov and V.~Usov for useful
comments and to N.~Kisilova,  Y.~Schukin and V.~Tsvetkova for
their help in translation of the text. The authors are extremely
grateful to reviewer for a kindly and qualified critique of this
work.}

\end{document}